
\input harvmac
%
%
%
%
\ifx\answ\bigans
\else
\output={
  \almostshipout{\leftline{\vbox{\pagebody\makefootline}}}\advancepageno
}
\fi
%
%
%

%
%

%
%
\def\UCSD#1#2{\noindent#1\hfill #2%
\bigskip\supereject\global\hsize=\hsbody%
\footline={\hss\tenrm\folio\hss}}
%
%
\def\abstract#1{\centerline{\bf Abstract}\nobreak\medskip\nobreak\par #1}
%
%
%
%
\edef\tfontsize{ scaled\magstep3}
 \tfontsize  \tfontsize
 \tfontsize \font\titlei=cmmi10 \tfontsize
\font\titleis=cmmi7 \tfontsize \font\titleiss=cmmi5 \tfontsize
\font\titlesy=cmsy10 \tfontsize \font\titlesys=cmsy7 \tfontsize
\font\titlesyss=cmsy5 \tfontsize  \tfontsize
\skewchar\titlei='177 \skewchar\titleis='177 \skewchar\titleiss='177
\skewchar\titlesy='60 \skewchar\titlesys='60 \skewchar\titlesyss='60
%
%
%
%
%
\def\inv{^{\raise.15ex\hbox{${\scriptscriptstyle -}$}\kern-.05em 1}}
\def\lbar{{\lower.35ex\hbox{$\mathchar'26$}\mkern-10mu\lambda}} 

%
%
%
%
\def\dsl{\,\raise.15ex\hbox{/}\mkern-13.5mu D} 
\def\delsl{\raise.15ex\hbox{/}\kern-.57em\partial}
\def\Ksl{\hbox{/\kern-.6000em\rm K}}
\def\Asl{\hbox{/\kern-.6500em \rm A}}
\def\Dsl{\hbox{/\kern-.6000em\rm D}} 
\def\Qsl{\hbox{/\kern-.6000em\rm Q}}
\def\gradsl{\hbox{/\kern-.6500em$\nabla$}}
%
%
\def\lspace{\ifx\answ\bigans{}\else\qquad\fi}
\def\lbspace{\ifx\answ\bigans{}\else\hskip-.2in\fi} 
%
%
\def\boxeqn#1{\vcenter{\vbox{\hrule\hbox{\vrule\kern3pt\vbox{\kern3pt
        \hbox{${\displaystyle #1}$}\kern3pt}\kern3pt\vrule}\hrule}}}
%
%
\def\mbox#1#2{\vcenter{\hrule \hbox{\vrule height#2in
\kern#1in \vrule} \hrule}}
%
%
%
%

  \def\CO{{\cal O}}

%
%
%
%
%

%

\def\bar#1{\overline{#1}}

\def\bra#1{\left\langle #1\right|}
\def\ket#1{\left| #1\right\rangle}

\def\darr#1{\raise1.5ex\hbox{$\leftrightarrow$}\mkern-16.5mu #1}

%
%
\def\frac#1#2{{\textstyle{#1\over #2}}} 
%
%
%
%

%
%
%
%

%
%
\def\ltap{\ \raise.3ex\hbox{$<$\kern-.75em\lower1ex\hbox{$\sim$}}\ }
\def\gtap{\ \raise.3ex\hbox{$>$\kern-.75em\lower1ex\hbox{$\sim$}}\ }
\def\gl{\ \raise.5ex\hbox{$>$}\kern-.8em\lower.5ex\hbox{$<$}\ }
\def\roughly#1{\raise.3ex\hbox{$#1$\kern-.75em\lower1ex\hbox{$\sim$}}}
%
%
        \def\etc{\hbox{\it etc.}}

\def\np#1#2#3{Nucl. Phys. B{#1} (#2) #3}
\def\pl#1#2#3{Phys. Lett. {#1}B (#2) #3}
\def\prl#1#2#3{Phys. Rev. Lett. {#1} (#2) #3}

\relax

\def\clebsch#1#2#3#4#5#6{\left(\left.
\matrix{{#1}&{#2}\cr{#4}&{#5}\cr}\right|\matrix{{#3}\cr{#6}}\right)}

\noblackbox
\def\lqcd{\Lambda_{\rm QCD}}
\def\nc{N_c}
\def\ln{large-$\nc$}
\def\Ln{Large-$\nc$}
\def\twobox#1#2{\vcenter{\hrule \hbox{\vrule height#2in
\kern#1in \vrule} \hrule \hbox{\vrule height#2in
\kern#1in \vrule}\hrule}}

\def\ga{g_A}

\centerline{{\titlefont{Baryon-Pion Couplings from Large-$\nc$ QCD}}}
\bigskip
\centerline{Roger Dashen and Aneesh V.~Manohar}
\smallskip
\centerline{{\sl Department of Physics, University of California at San
Diego, La Jolla, CA 92093}}
\bigskip
\vfill
\abstract{We derive a set of consistency conditions for the pion-baryon
coupling constants in the \ln\ limit of QCD. The consistency conditions have a
unique solution which are precisely the values for the pion-baryon coupling
constants in the Skyrme model. We also prove that non-relativistic $SU(2N_f)$
spin-flavor symmetry (where $N_f$ is the number of light flavors) is a
symmetry of the baryon-pion couplings in the \ln\ limit of QCD. The
symmetry breaking corrections to the pion-baryon couplings vanish to
first order in $1/\nc$. Consistency conditions for other couplings, such
as the magnetic moments are also derived.
}
\vfill
\UCSD{\vbox{\hbox{UCSD/PTH
93-16}\hbox{hep-ph/9307241}}}{July 1993}


In this paper, we will study the baryon-pion coupling constants in the \ln\
limit~\ref\thooft{G. 't Hooft, \np{72}{1974}{461}} . We derive a set of
consistency conditions for the baryon-pion couplings
that must be satisfied by \ln\ QCD. These consistency conditions completely
determine all the pion-baryon coupling constants in terms of one overall
normalization constant. The consistency conditions require that there be an
infinite tower of degenerate baryon states with $I=J$. The ratios of the
pion-baryon couplings of this baryon tower are precisely those given by the
Skyrme model~\ref\anw{T.H.R. Skyrme, Proc. R. Soc. London A260 (1961)
127\semi G.S.~Adkins, C.R.~Nappi, and E.~Witten, \np{228}{1983}{552}\semi
E. Witten, \np{B223}{1983}{433}},
or by the non-relativistic quark model~\ref\nrqm{G. Karl and J. Paton,
Phys.~Rev.~D30 (1984) 238}, which is known to be
equivalent to the Skyrme model in the \ln\ limit~\ref\am{A.V.~Manohar,
\np{248}{1984}{19}}. This implies that the
\ln\ limit of QCD has a $SU(2N_f)$ spin-flavor symmetry, where $N_f$ is the
number of light flavors. This result is unusual in that the $SU(2N_f)$
symmetry is only a symmetry of the baryon sector of \ln\ QCD, but not of the
meson sector. After this work was completed, we discovered that the
first consistency condition proved in this paper using pion-nucleon
scattering was derived earlier using the same method by Gervais and
Sakita \ref\gs{J.-L. Gervais and B.~Sakita, \prl{52}{1984}{87},
Phys.~Rev.~D30 (1984) 1795}. They also realized that there is a
contracted $SU(4)$ algebra in the \ln\ limit of QCD.
We believe that the other consistency conditions derived in this paper, the
connection with chiral perturbation theory, and the study of $1/\nc$
corrections
(which will be presented elsewhere~\ref\dm{R. Dashen and A.V. Manohar, UCSD/PTH
93-18}\ref\djm{R. Dashen, E. Jenkins and A.V. Manohar, UCSD/PTH 93-21}) are
new.

The \ln\ counting rules for meson-baryon scattering were analysed by
Witten~\ref\witten{E. Witten, \np{160}{1979}{57}}, who showed that
meson-baryon  scattering
amplitudes at fixed energy must be of order one. The incident pion can
couple to
any of the quarks in the baryon with an amplitude $1/\sqrt\nc$. If the
emitted pion couples
to the same quark as the incident pion, there are $\nc$ choices for the
quark, so the net amplitude is order one. If the emitted pion couples to a
different quark, then there are $\nc^2$ ways to choose the two quarks, but
there must also be at least a single gluon exchange between the two quarks to
transfer energy from the incident pion to the final pion. The gluon produces a
suppression of $1/\nc$, so again, the amplitude is of order one.
The \ln\ counting rules show that the axial vector coupling constant $\ga$ of a
baryon is of order $\nc$, since the axial current can be inserted on any of the
$\nc$ quark lines. It is also easy to see that the pion decay constant $f_\pi$
is of order $\sqrt{\nc}$. The pion-baryon vertex is proportional to
$g_A{\bf q}/f_\pi$, where ${\bf q}$ is the pion momentum,
and grows as $\sqrt{\nc}$ at fixed pion energy. The consistency
conditions follow from this simple result.

Let us assume that the \ln\ QCD baryon spectrum contains a state with
$I=J=1/2$, which we will call the nucleon. The nucleon
is infinitely heavy, and can be treated as a static fermion interacting with
the pion~\ref\jm{E.~Jenkins and A.V.~Manohar, \pl{255}{1991}{558}\semi
S.~Weinberg, \pl{251}{1990}{288}}.
The pion-nucleon scattering amplitude is dominated by the graphs of
\fig\piNfig{Diagrams contributing to $\pi N$ scattering. Fig 1(c) is suppressed
by $1/\nc$ relative to figs.~1(a) and 1(b).}. The axial current matrix element
in
the nucleon can be written as \eqn\axialmatrix{
\bra{N} \bar\psi\gamma^i\gamma_5 \tau^a\psi \ket{N} = g \nc \bra{N} X^{ia}
\ket{N},
}
where $\bra{N} X^{ia} \ket{N}$ and $g$ are of order one. The coupling
constant $g$ has been factored out so that the normalization of $X^{ia}$ can
be chosen conveniently. $X^{ia}$ is an operator (or
$4\times4$ matrix) defined on nucleon states which has a finite \ln\ limit.
The pion-nucleon scattering amplitude for $\pi^a(q) + N (k) \rightarrow
\pi^b(q^\prime)+ N(k^\prime)$ is (neglecting the third graph which is
suppressed by $1/\nc^2$)
\eqn\amp{
-i\ q^i q^{\prime\,j} {\nc^2g^2\over f_\pi^2}
\left[{ 1\over q^0} X^{jb }X^{ia}
- {1 \over q^{\prime\,0} } X^{ia}X^{jb}\right],
}
where the amplitude is written in the form of an operator acting on nucleon
states. Both initial and final nucleons are on-shell, so $q^0=q^{\prime\,0}$.
The product of the $X$'s in eq.~\amp\ then sums over the possible spins and
isospins of the intermediate nucleon. Since $f_\pi\sim\sqrt\nc$, the overall
amplitude is of order $\nc$, which violates unitarity, and also contradicts the
\ln\ counting rules of Witten. Thus \ln\ QCD with a $I=J=1/2$ nucleon multiplet
interacting with a pion is an inconsistent field theory. There must be
other states that cancel the order $\nc$ amplitude in eq.~\amp\ so that
the total amplitude is order
one, and consistent with unitarity. One can then generalize $X^{ai}$ to be an
operator on this degenerate set of baryon states, with matrix elements equal to
the corresponding axial current matrix elements. With this generalization, the
form of eq.~\amp\ is unchanged. Thus we obtain the first consistency condition
for QCD,
\eqn\consi{
\left[X^{ia},X^{jb}\right]=0,
}
so that the axial currents are represented by a set of
operators $X^{ia}$ that commute in the \ln\ limit. The solution of the
constraint is non-trivial, because we also have the commutation
relations
\eqn\xjcomm{
\left[J^i, X^{jb}\right]=i\,\epsilon_{ijk}\, X^{kb},\qquad
\left[I^a, X^{jb}\right]=i\,\epsilon_{abc}\, X^{jc},
$$
$$
\left[J^i,J^j\right]=i\,\epsilon_{ijk}\, J^k,\qquad
\left[I^a,I^b\right]=i\,\epsilon_{abc}\, I^c,\qquad
\left[I^a,J^i\right]=0,
}
since $X^{ia}$ has spin one and isospin one. It is simple to prove that
eqs.~\consi\ and \xjcomm\ form a non-semisimple Lie Algebra, and have no
non-trivial finite dimensional representations~\djm.

The solutions of the consistency condition eq.~\consi\ requires that there be
an infinite tower of degenerate baryon states, and also determines the ratios
of the pion-baryon couplings between the states. We will make one simplifying
assumption in this letter; the degnerate baryon spectrum consists only of
states $I=J=1/2,3/2,\ldots$, where the sequence can be finite or
infinite. (This assumption is not neccessary \djm.)
The states will be denoted by $\ket{J,J_3,I_3}$. The reduced matrix
element of $X^{ai}$ between baryon states can be written as
\eqn\reduced{
\bra{J^\prime,m^\prime,\alpha^\prime} X^{ia} \ket{J,m,\alpha}
= X(J,J^\prime) \sqrt{{2J+1\over 2J^\prime+1}} \clebsch J 1 {J^\prime} m i
{m^\prime} \clebsch J 1 {J^\prime} \alpha a {\alpha^\prime},
}
where $X(J,J^\prime)$ is a reduced matrix element. The normalization
constant has been chosen so that $X(J,J^\prime)=X(J^\prime,J)$. Since
$X^{ia}$ is a tensor with spin one and isospin one, it can only couple
states with $\Delta J=0, \pm 1$, and the independent
reduced matrix elements are $X(J,J)$ and $X(J,J+1)$.
Taking the matrix element of  eq.~\consi\ between $\ket{J,m,\alpha}$ and
$\ket{J^\prime,m^\prime,\alpha^\prime}$, and inserting a complete set of
intermediate states gives
\eqn\explicit{
0=\sum_{J_1 m_1 \alpha_1}
\bra{J^\prime,m^\prime,\alpha^\prime}
X^{ia} \ket{J_1,m_1,\alpha_1}
\bra{J_1,m_1,\alpha_1} X^{jb}\ket{J,m,\alpha}
 - \left(ia\leftrightarrow jb\right).
}
We will first show that eq.~\explicit\ has a unique solution
Consider the case $J^\prime=J+1$, where only intermediate states
$J_1=J,J+1$ contribute. The condition in eq.~\explicit\ reduces to
\eqn\outlinei{
0 \sim X(J,J)\, X(J,J+1) + X(J,J+1)\, X(J+1,J+1),
}
where we have only shown the structure of the reduced matrix elements
that contribute, and neglected all numerical factors, signs, \etc\
Similarly, the diagonal matrix element $J^\prime=J$ of
eq.~\explicit\ has the form
\eqn\outlineii{
0 \sim X(J,J-1)^2 + X(J,J)^2 + X(J,J+1)^2.
}
If we are given $X(J,J)$ and $X(J-1,J)$, eq.~\outlineii\ determines
$X(J,J)$ and eq.~\outlinei\ determines $X(J+1,J+1)$, so that
all the reduced matrix elements are uniquely determined by recursion starting
from the lowest ones. If we assume that the lowest state is $J=1/2$,
all the reduced matrix elements are determined in terms of
$X(1/2,1/2)$, since $X(1/2,-1/2)=0$. Since eq.~\explicit\ is  homogeneous
in $X(J,J^\prime)$, the starting value $X(1/2,1/2)$ can be set equal to
one, which determines the solution uniquely up to an overall
normalization constant. To construct eq.~\outlinei\ and \outlineii\
explicitly, substitute eq.~\reduced\ into eq.~\explicit, and project
onto the pion-nucleon scattering channel with spin $H$ and isospin $K$ by
multiplying the equation by
$$
\clebsch J 1 H m i h \clebsch J 1 K \alpha a \eta
\clebsch {J^\prime} 1 {H^\prime} {m^\prime} j {h^\prime}
\clebsch {J^\prime} 1 {K^\prime} {\alpha^\prime} b {\eta^\prime}.
$$
The resulting equation for the reduced matrix elements can be written
as
\eqn\outlineiii{\eqalign{
X(J,H) X(J^\prime,H)& \delta_{HK} = \cr(2H+1) \sum_{J_1} & (2J_1+1)
\left\{\matrix{1&J_1&J\cr1&H&J^\prime\cr}\right\}
\left\{\matrix{1&J_1&J\cr1&K&J^\prime\cr}\right\}
X(J,J_1) X(J^\prime,J_1).
}}
The solution to this equation has been shown to be unique, so it is
sufficient to verify that $X(J,J)=X(J,J+1)=1$ for all $J$ is a solution
to this set of equations, using the symmetry properties of the
$6j$-symbols and the identity~\ref\edmonds{A.R.~Edmonds, Angular Momentum
in Quantum Mechanics, Princeton University Press, Princeton 1974}
\eqn\sixjiden{
\sum_K (2K+1)(2H+1)
\left\{\matrix{J_1 &J_2 &H\cr J_3 &J_4 &K}\right\}
\left\{\matrix{J_1 &J_2 &L\cr J_3 &J_4 &K}\right\} = \delta_{HL}.
}
A similar result can also be proved for the infinite tower
$I=J=0,1,\ldots$ if one starts from the lowest state $I=J=0$. The two
possible towers correspond to the \ln\ limit of QCD with $\nc$ odd and
even, respectively. The solution eq.~\reduced\
for the pion-baryon couplings in \ln\ QCD is identical to the results
obtained in the Skyrme model and the non-relativistic quark model in the
\ln\ limit.

The algebra in eq.~\consi\ and \xjcomm\ is a contracted SU(4) algebra.
Consider the embedding $SU(4)\rightarrow SU(2)\otimes SU(2)$ where
$4\rightarrow (2,2)$. If the generators of $SU(2)\otimes SU(2)$ in
the defining representation are
$I^a$, and $J^i$, the $SU(4)$ generators in the defining representation
are $J^i\otimes 1$, $1\otimes I^a$ and $J^i\otimes I^a$, which we will
call $I^a$, $J^i$ and $G^{ia}$ respectively. (The properly normalized
$SU(4)$ generators are $I^a/\sqrt{2}$, $J^i/\sqrt{2}$ and $\sqrt2\,
G^{ia}$.)
The algebra of \ln\ QCD is given by taking the limit
\eqn\xlimit{
X^{ia} = \lim_{\nc\rightarrow\infty} {G^{ia} \over \nc},
}
(up to an overall normalization of the $X^{ia}$). The commutation
relations of $SU(4)$,
\eqn\sufour{
\eqalign{
&\left[J^i,J^j\right]=i\,\epsilon_{ijk}\,J^k,\cr
&\left[I^a,G^{ia}\right] = i\,\epsilon_{abc}\, G^{jc},\cr
&\left[I^a,J^i\right]=0,}
\qquad\eqalign{
&\left[I^a,I^b\right]=i\,\epsilon_{abc}\, I^c,\cr
&\left[J^i,G^{jb}\right] = i\,\epsilon_{ijk}\, G^{kb},\cr
&\left[G^{ia},G^{jb}\right] = \frac i 4\, \epsilon_{ijk} \delta_{ab}\, J^k +
\frac i 4\,\epsilon_{abc} \delta_{ij}\, I^c,
}}
turn into the commutation relations eqs.~\consi--\xjcomm\ in the \ln\ limit.
The non-trivial irreducible representations of the contracted algebra
are obtained by taking the limit of $SU(4)$ representations for which
$G^{ia}$ is of order $\nc$, so that $X^{ia}$ is finite. We have
constructed the simplest such representation explicitly above, and it
corresponds to taking the $\nc\rightarrow\infty$ limit of the completely
symmetric $SU(4)$ tensor with $\nc$ boxes. Thus we see that the \ln\
limit of QCD has a contracted $SU(4)$ symmetry in the baryon sector.
This is sufficient to prove that the predictions for the pion-baryon
couplings are identical to those of the Skyrme model or non-relativistic
quark model, and exhibit $SU(4)$ symmetry. This explains the origin of
the non-relativistic $SU(4)$
symmetry of baryons in QCD.

There is an alternative approach to deriving consistency conditions
for \ln\ QCD that leads to a condition equivalent to eq.~\consi\ as well
as to a second consistency condition. Consider the one-loop radiative
corrections in the baryon sector, given by the diagrams of
\fig\oneloopfigs{Diagrams contributing to the one-loop corrections of the
pion-baryon couplings.}. The one loop renormalization of the pion-baryon
coupling
constants is proportional to \eqn\pbshift{
\delta \left(g N_c X^{ia}\right) \sim {g^2 N_c^2\over 16 \pi^2 f_\pi^2}
\left[ X^{jb},\left[X^{jb},X^{ia}\right]\right]
m_\pi^2 \log m_\pi^2/\mu^2,
}
so that a term of order $N_c$ gets a correction of order $N_c^2$. The
pion mass is a free parameter of the theory, and is of order $\sqrt{\lqcd
m_q}$ and finite in the \ln\ limit. Thus the stability of the pion-baryon
couplings under an infinitesimal shift in the light quark mass, or
equivalently,
stability under radiative corrections gives the consistency condition
\eqn\consii{
\left[ X^{jb},\left[X^{jb},X^{ia}\right]\right]=0,
}
which is a cubic relation between the $X$'s. One can show that this
consistency condition also leads to a set of recursion relations for the
reduced matrix elements of $X$, with a unique solution that is the same
as that obtained earlier. The details will be given elsewhere.
Thus one-loop chiral perturbation theory in the \ln\ limit implies that
there must be an infinite tower of degenerate baryon states of the form
$I=J=1/2,3/2,\ldots$, with pion-baryon couplings which are precisely
those given by the Skyrme model. In addition, the one-loop corrections
must cancel exactly at leading order in $\nc$.

The quark counting rules of Witten do not lead to any violations of
unitarity or consistency conditions in the \ln\ limit. In a quark
picture, one sums over all possible intermediate quark states, which is
equivalent to summing over all intermediate baryon states. The singular
\ln\ behavior of the pion-baryon scattering amplitude or of one-loop
chiral perturbation theory arises because the intermediate state was
projected onto a single baryon state, the nucleon. It is this projection
that leads to an inconsistency in the \ln\ limit. \Ln\ QCD requires the
existence of an infinite tower of degenerate resonances which must be
included as intermediate states for chiral perturbation theory to be
consistent. The loop graphs cancel exactly to leading order as
$\nc\rightarrow\infty$. It was recently
suggested~\ref\jmdelta{E.~Jenkins and A.V.~Manohar, \pl{259}{1991}{353}}
that $\Delta$ resonances must be included as
intermediate states in baryon chiral perturbation theory for a
consistent expansion, and that there is a large cancellation between
intermediate $\Delta$ and nucleon states. The computation was done for
$N_c=3$ and the physical values of the $\pi NN$, $\pi N\Delta$ and
$\pi\Delta\Delta$ couplings.
It was also found that the best fit values
for these couplings were remarkably close to the values obtained using
non-relativistic $SU(6)$ symmetry, and that the large cancellation
occured only for values of the pion-couplings near the $SU(6)$ values,
but not for generic couplings. It was suggested that the baryon-pion
couplings should have a non-relativistic $SU(6)$
symmetry~\ref\hungary{E.~Jenkins and A.V.~Manohar, {\sl Baryon Chiral
Perturbation Theory}, in Effective Field Theories of the Standard Model, ed. by
U. Meissner, World Scientific (1992)}. These results are exact in the \ln\
limit
of QCD. We will also show that the leading corrections to the $SU(6)$ relations
are $1/\nc^2$ \dm, which explains why the values of the pion-baryon couplings
observed experimentally are so close to their $SU(6)$ values.

The wave-function renormalization graph also gives a
mass shift of the baryons proportional to
\eqn\mshift{
{g^2 N_c^2\over 16 \pi f^2}  X^{ia} X^{ia} m_\pi^3 .
}
This mass shift cannot break the degeneracy of the baryon spectrum, or
else the pion-baryon coupling consistency equations eq.~\consi\ or
\consii\ cannot be satisfied. Thus we require that
\eqn\consiii{
X^{ia} X^{ia} = C_X + \CO\left({1\over \nc^2}\right),
}
where $C_X$ is a constant. This is the third consistency condition that
must be satisfied by \ln\ QCD. It is a trivial check to see that the
solution eq.~\reduced\ with $X(J,J^\prime)=1$ satisfies this consistency
condition, with $C_X=3$.

The consistency conditions eq.~\consi\ and \consii\ are equivalent if
one also assumes eq.~\consiii. Eq.~\consii\ can be written in the form
$$\eqalign{
0=\left[ X^{jb},\left[X^{jb},X^{ia}\right]\right]&= X^{jb} X^{jb} X^{ia}
+ X^{ia} X^{jb} X^{jb} - 2 X^{ia} X^{jb} X^{ia}, \cr
&= 2 C_X X^{ia} -  2 X^{ia} X^{jb} X^{ia},
}$$
which leads to an equivalent form for eq.~\consii\
\eqn\consiv{
X^{jb} X^{ia} X^{jb} = C_X X^{ia}.
}
Now assume eq.~\consiii\ and eq.~\consiv, and define
$M^{ia,jb} = i\left[X^{ia},X^{jb}\right]$. Then
$$\eqalign{
\sum_{ia,jb} \left[M^{ia,jb}\right]^2 = \sum_{ia,jb}
&X^{ia}X^{jb}X^{jb} X^{ia} + X^{jb} X^{ia}X^{ia}X^{jb}\cr
&\qquad- X^{jb}X^{ia}X^{jb}X^{ia} - X^{ia}X^{jb}X^{ia}X^{jb}=0,
}$$
using eq.~\consiii\ and eq.~\consiv. But $M^{ia,jb}$ is a Hermitian
operator, and so $\sum_{ia,jb} \left[M^{ia,jb}\right]^2=0$ implies that
$M^{ia,jb}=0$ for all $ia$ and $jb$. This proves eq.~\consi. To prove
the converse, assume eq.~\consi\ and eq.~\consiii. Then eq.~\consiv\
follows trivially by multiplying eq.~\consi\ by $X^{jb}$ and summing
over $jb$.

We can also look at other operators in the \ln\ limit of QCD
such as magnetic moments, or mass splittings. These operators can be
treated as a vertex operator $V^{\ldots}$, where the indices on
$V^{\ldots}$ depend on its spin and isospin transformation properties.
The stability of insertions
of the vertex operator to one-loop chiral  corrections implies the
consistency condition
\eqn\consv{
\left[X^{ia},\left[X^{ia},V^{\ldots}\right]\right]=0
}
which can be written using eq.~\consiii\ as
\eqn\consvi{
X^{ia}V X^{ia} = C_X V.
}
The operator $=\left(X^{ia}\right)_{\alpha\beta}
\left(X^{ia}\right)_{\lambda\tau}$ is a linear operator, and
eq.~\consvi\ implies that allowed vertex operators in the \ln\ limit
must be eigenstates of the Casimir operator $X^{ia}X^{ia}$ with
eigenvalue $C_X$. An example
is the isovector magnetic moment vertex $\mu_V$. It must satisfy
eq.~\consvi\ and has the same spin and isopsin as the pion vertex $X$.
Therefore, $\mu_V^{ia}$ must be a constant times $X^{ia}$, as in the
Skyrme model. The same result can be derived by looking at the $\nc$
dependence of pion photoproduction. The scattering formalism also provides a
somewhat stronger version of eq.~\consv\ when $V$ is an
isospin-splitting mass insertion
$\Delta M$. Expanding the energy denominators $q^0$ and $q^{\prime\,0}$
in eq.~\amp\ and generalizing to scattering from an arbitrary bayon
yields
\eqn\consvii{
\left[X^{ia},\left[X^{jb},\Delta M\right]\right]=0,
}
with an error of order $1/\nc$. Unlike the axial couplings, the single
commutator $\left[X^{ia}, \Delta M\right]$ need not vanish.

The results presented here can be generalized to an arbitrary number of
flavors $N_f$. The consistency conditions eqs.~\consi\consiii\consvii\ are
still valid, and prove that \ln\ QCD has a contracted $SU(2N_f)$
symmetry. The recursion relations such as eq.~\outlineiii\ are more
complicated, because the isospin $6j$-symbol is replaced by the flavor
$6j$-symbol for $SU(N_f)$.

In conclusion, we have derived a set of consistency conditions that must
be satisfied by \ln\ QCD, and have proved that \ln\ QCD has a contracted
$SU(4)$ symmetry (for $N_f=2$). The solutions of the consistency
equations lead to a
unique (minimal) solution for the pion-baryon coupling constants, which are
identical to those of the Skyrme model or non-relativistic quark model,
and also to the existence of an infinite tower of baryon resonances. The
$1/\nc$ corrections to the pion-baryon coupling constant ratios can be
shown to vanish \dm. We have also briefly discussed
consistency conditions for other operators, and the generalization to an
arbitrary number of flavors. The details will be presented elsewhere.
The methods used here have been used to
confirm the soliton picture of baryons containing a heavy
quark~\ref\jenkins{E.~Jenkins, UCSD/PTH 93-17}\ in which the baryon is
treated as a bound state of a
soliton and a heavy meson~\ref\callan{C. Callan and I. Klebanov,
\np{262}{1985}{365}\semi
C. Callan, K. Hornbostel, and I. Klebanov, \pl{202}{1988}{260}}\ref\jmw{
E.~Jenkins, A.V.~Manohar, and M.B.~Wise, \np{396}{1993}{27}\semi
Z.~Guralnik, M.~Luke and A.V.~Manohar, \np{390}{1993}{474}\semi
E.~Jenkins and A.V.~Manohar, \pl{294}{1992}{273}\semi
E.~Jenkins, M.B.~Wise and A.V.~Manohar,
\np{396}{1993}{38}}\ref\mrho{M.~Rho, D.O.~Riska and N.N.~Scoccola,
\pl{251}{1990}{597}, Z.~Phys.~A341 (1992) 343\semi
Y.~Oh, D.~Min, M.~Rho and N.N.~Scoccola, Nucl.~Phys.~A534 (1991) 493},
and to prove that the leading correction to the baryon mass
splittings is proportional to ${\bf J}^2$~\ref\ejii{E. Jenkins, UCSD/PTH
93-19}.

We would like to thank E.~Jenkins for helpful discussions. This work was
supported in part by DOE grant DOE-FG03-90ER40546 and by a PYI award
PHY-8958081.
\listrefs \listfigs
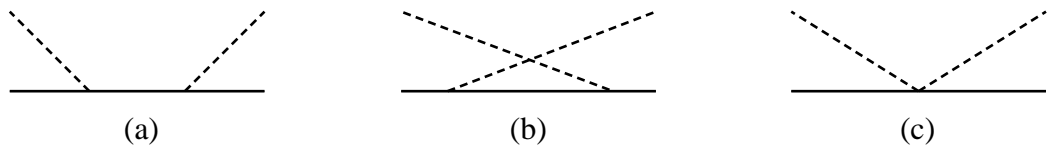
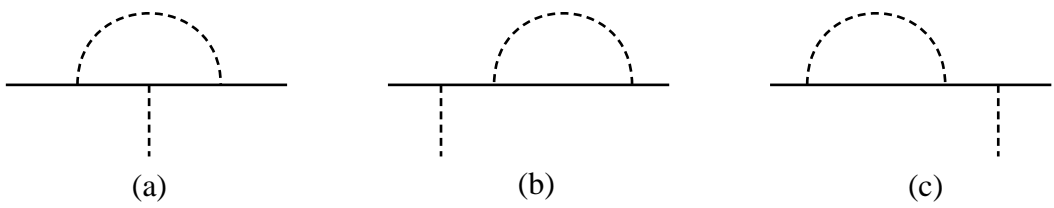
\bye